\documentclass[twocolumn,superscriptaddress,prl]{revtex4}   % twocolumn %,showpacs
\usepackage{graphicx,amsmath,mathrsfs}
\usepackage{amssymb}
\usepackage{amsthm,multirow}
\usepackage{bm,bbm}
\usepackage{float}

\usepackage{subfigure}   

\usepackage{color}

\def\bc{\begin{center}}
\def\ec{\end{center}}

\newcommand{\up}{\uparrow}
\newcommand{\dw}{\downarrow}
\newcommand{\pd}{{\phantom{\dagger}}}

\begin{document}
%%%%%%%%%%%%%%%%%%%%%%%%%%%%%%%%%%%%%%%%%%%%%
\title{Quantum Engineering of Majorana Fermions}
%%%%%%%%%%%%%%%%%%%%%%%%%%%%%%%%%%%%%%%%%%%%%
\author{Eric Mascot}
\affiliation{University of Illinois at Chicago, Chicago, IL 60607, USA}
\author{Sagen Cocklin}
\affiliation{University of Illinois at Chicago, Chicago, IL 60607, USA}

\author{Stephan Rachel}
\affiliation{School of Physics, University of Melbourne, Parkville, VIC 3010, Australia}
\author{Dirk K.\ Morr}
\affiliation{University of Illinois at Chicago, Chicago, IL 60607, USA}
%%%%%%%%%%%%%%%%%%%%%%%%%%%%%%%%%%%%%%%%%%%%%
\date{\today}

\maketitle

%%%%%%%%%%%%%%%%%%%%%%%%%%%%%%%%%%%%%%%%%%%%%
%
%                                                       A B S T R A C T
%
%%%%%%%%%%%%%%%%%%%%%%%%%%%%%%%%%%%%%%%%%%%%%

\textbf{
Chiral superconductors have the ability to host topologically protected Majorana zero modes which have been proposed as future qubits for topological quantum computing. The recently introduced magnet--superconductor hybrid (MSH) systems consisting of magnetic adatoms deposited on the surface of conventional $s$-wave superconductors represent a promising candidate for the creation, detection, and manipulation of Majorana modes via scanning tunneling microscopy techniques. Here, we present four examples demonstrating the ability to engineer Majorana fermions in nanoscopic MSH systems by using atomic manipulation techniques  to change the system's shape or magnetic structure. This allows for the dimensional tuning of Majorana modes between one and two dimensions, the identification of the system's topological invariant -- the Chern number -- in real space, the creation of chiral Majorana modes of arbitrary length and shape along magnetic domain boundaries, and the creation of a topological switch using magnetic skyrmions. These examples of Majorana fermion engineering can be realised with current experimental techniques and will promise new avenues for the first prototypes of Majorana-based quantum devices.}

\vspace{5pt}

The recent observations of Majorana modes in one- \,\cite{mourik-12s1003,nadj-perge-14s602,ruby-15prl197204,pawlak-15npjqi16035,kim-18sa5251} and two-dimensional \,\cite{menard16-nc2040,palacio-morales18arXiv} topological superconductors hold the promise for topology-based technologies and topological quantum computation \,\cite{nayak-08rmp1083}. However, the realization of these technologies will not only require the ability to engineer Majorana fermions in nanoscale system, but also to manipulate them spatially at the length scale of a few lattice constants. An important step towards the first objective was recently taken via single-atom manipulation of MSH structures, where chains of magnetic Fe adatoms were built atom-by-atom on a superconducting Re surface using the tip of a scanning tunneling microsocope\,\cite{kim-18sa5251}, allowing to visualize the emergence of Majorana bound states. Similarly, interface engineering in MSH structures has proven crucial in the creation of two-dimensional topological superconductivity and the direct visualization of chiral Majorana edge modes \,\cite{palacio-morales18arXiv}. Another opportunity for the manipulation of MSH structures arises from the ability to imprint complex magnetic structures at the nanoscale: either by writing and deleting skyrmions \cite{romming-15prl177203,romming-13s636} using scanning tunneling microscopy (STM) techniques, or by engineering magnetic domain walls \cite{Yasuda1311} via atomic force microscopy (AFM). It is the combination of all of these techniques that allow for the simultaneous engineering of real space and magnetic structures which likely holds the key to the realization of topological quantum devices.

To investigate the engineering of Majorana fermions, we consider magnetic-superconducting hybrid structures in which magnetic adatoms are placed on the surface of a conventional $s$-wave superconductor with a Rashba spin-orbit interaction. Such a system is described by the Hamiltonian
\begin{equation}\begin{split}
&H_{\rm MSH} = -t \sum_{{\langle \bf r r^\prime}\rangle, \sigma} \left( c_{{\bf r},\sigma}^\dag c^\pd_{{\bf r^\prime},\sigma}
+ {\rm H.c.}\right)  - \mu \sum_{{\bf r},\sigma} c_{{\bf r},\sigma}^\dag c_{{\bf r},\sigma}
\\
&\quad+\,i \alpha  \sum_{{\bf r},\alpha \beta} c_{{\bf r},\alpha}^\dag \left( {\bm {\hat \delta}} \times {\bm \sigma}\right)^z_{\alpha \beta} c^\pd_{{\bf r}+{\hat \delta},\beta}\\
&\quad+ \,J {\bf S} \cdot \sum_{{\bf R},\sigma,\sigma'} c_{{\bf R},\sigma}^\dag {\bm \sigma}_{\sigma\sigma'} c^\pd_{{\bf R},\sigma'}
+ \Delta \sum_{\bf r} \left( c_{{\bf r},\up}^\dag c^\dag_{{\bf r},\dw} + {\rm H.c.} \right)
%\notag \\
%-&t_{\rm tip} \sum_{\sigma} \left( c_{{\bf r},\sigma}^\dag d^\pd_{\sigma} + {\rm H.c.} \right)\ , %  + H_{\rm tip}\ ,
\label{ham-realspace}
\end{split}
\end{equation}
where $c_{{\bf r},\alpha}^\dag$ creates an electron at lattice site ${\bf r}$ with spin $\alpha$, and ${\bm \sigma}$ is the vector of spin Pauli matrices. For concreteness, we consider a square lattice, with $-t$ being the hopping amplitude between nearest-neighbor sites and $\mu$ being the chemical potential; we emphasize, however, that all results remain true for other lattice geometries as recently demonstrated \,\cite{RachelMorr17,palacio-morales18arXiv} and expected for topologically non-trivial matter.
Moreover, $\alpha$ denotes the Rashba spin-orbit coupling arising from the breaking of the inversion symmetry at the surface\,\cite{nadj-perge-14s602} with $\bm \delta$ being the vector connecting nearest neighbor sites, and $J$ is the magnetic exchange coupling of a magnetic defect located at ${\bf R}$. Due to the full superconducting gap, which suppresses Kondo screening, we consider the magnetic moments to be static in nature. Below, we provide four examples of how engineering of Majorana modes in nanoscale MSH structures can be achieved through the manipulation of their real space shape and magnetic structure.

%%%%%%%%%%%%%%%%%%%%%%%%%%%%%%%%%%%%%%%%%%%%%
%                  F I G .     1
%%%%%%%%%%%%%%%%%%%%%%%%%%%%%%%%%%%%%%%%%%%%%
\begin{figure*}[t]
\centering
\includegraphics[width=15.2cm]{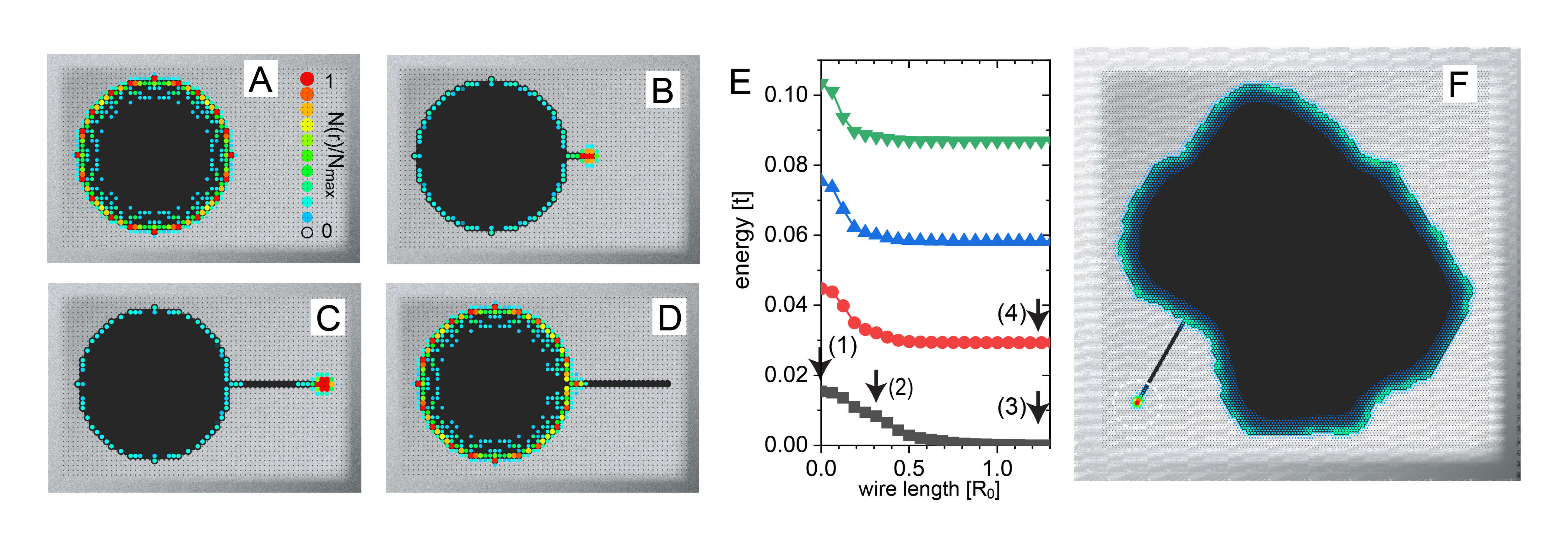}\\
\caption{ \textbf{Dimensional Tuning of Majorana Modes} MSH hybrid structure consisting of a Shiba island of radius $R=16 a_0$ (with $C=-1$) and chain with parameters $(\mu,\Delta,\alpha,J)=(-4t, 1.2t, 0.45t, 2.6t)$ (black circles denote sites with magnetic adatoms.). {\bf (A)}-{\bf (C)} LDOS of the lowest energy Majorana mode with increasing chain length, corresponding to arrows (1)-(3) in {\bf (E)}. {\bf (D)}  LDOS of the second lowest energy Majorana mode, corresponding to arrow (4) in {\bf (E)}. {\bf (E)} Evolution of the 4 lowest energy levels with increasing chain length, units of the diameter, $R_0$, of the Shiba island.{\bf (F)} Zero-energy LDOS of the experimentally realized Fe island \,\cite{palacio-morales18arXiv} placed with a Shiba chain attached to it. Dashed white circle indicates the location of the Majorana bound state at the end of the chain.}
\label{fig:fig1}
\end{figure*}
%%%%%%%%%%%%%%%%%%%%%%%%%%%%%%%%%%%%%%%%%%%%%
The recent experimental observations of Majorana fermions at the end of one-dimensional (1D) chains \cite{nadj-perge-14s602,ruby-15prl197204,pawlak-15npjqi16035,kim-18sa5251} and at the edges of two-dimensional (2D) islands \,\cite{menard16-nc2040,palacio-morales18arXiv} of magnetic adatoms, referred to as Shiba chains and islands, respectively, have raised the intriguing question of whether it is possible to adiabatically tune between topological phases in 1D and 2D MSH structures without undergoing a phase transition. This question is of particular interest not only because topological superconductors in 1D and 2D are in different homotopy groups -- with homotopy group $Z_2$ in 1D, and $Z$ in 2D \,\cite{schnyder-08prb195125,kitaev09} --, but also because advances in atomic manipulation techniques have rendered the experimental realization of such tuning possible \cite{kim-18sa5251}. To investigate this possibility, we consider a system in which both the 1D chains and 2D islands of magnetic adatoms induce topological superconductivity, with Chern number $C=1$ in the latter case. To interpolate between Shiba chains and islands, we attach an increasingly longer chain of magnetic adatoms to a Shiba island, as shown in Figs.~\ref{fig:fig1}A-C. With no chain present, the Shiba island possesses a chiral Majorana mode that is localized near the edge of the island, and forms a dispersing, 1D edge mode that traverses the superconducting gap. Each chiral edge mode is comprised of two Majorana fermions, which for the lowest energy mode are located at small, but finite energy $E=\pm \epsilon$ [Fig.~\ref{fig:fig1}E] (this non-zero energy, and the discreteness of the modes arises from the finite size of the island \cite{RachelMorr17}). The LDOS of the lowest energy mode is shown in Fig.~\ref{fig:fig1}A. When a chain is attached to the island, and its length is increased,  spectral weight is transferred from the island's lowest energy Majorana edge mode to the end of the chain [Fig.~\ref{fig:fig1}B]. When the chain is sufficiently long [Fig.~\ref{fig:fig1}C] it possesses a localized Majorana fermion at its end, while a second Majorana fermion remains delocalized along the edge of the island. Note that the spatially integrated spectral weight of the zero-energy state is exactly split between the dispersive Majorana edge fermion and the bound state at the end of the chain. Concomitant with the increasing separation between these two Majorana fermions, the energy of the lowest energy states decreases smoothly and monotonically [Fig.~\ref{fig:fig1}E], implying that the system remains in a topological phase throughout this evolution, i.e., without undergoing a phase transition. At the same time, the higher energy Majorana modes remain entirely localized along the edge of the island, as shown in Fig.~\ref{fig:fig1}D for the second lowest energy state. This example demonstrates that it is not only possible to adiabatically tune between 1D and 2D topological superconductivity via atomic manipulation (and hence spatially separate Majorana fermions without the creation of magnetic vortices \cite{alicea-12rpp076501}), but also to design a single system exhibiting both localized and dispersive Majorana zero modes.
%
%%%%%%%%%%%%%%%%%%%%%%%%%%%%%%%%%%%%%%%%%%%%%
%                  F I G .     2
%%%%%%%%%%%%%%%%%%%%%%%%%%%%%%%%%%%%%%%%%%%%%
\begin{figure}[t!]
\centering
\includegraphics[width=8.2cm]{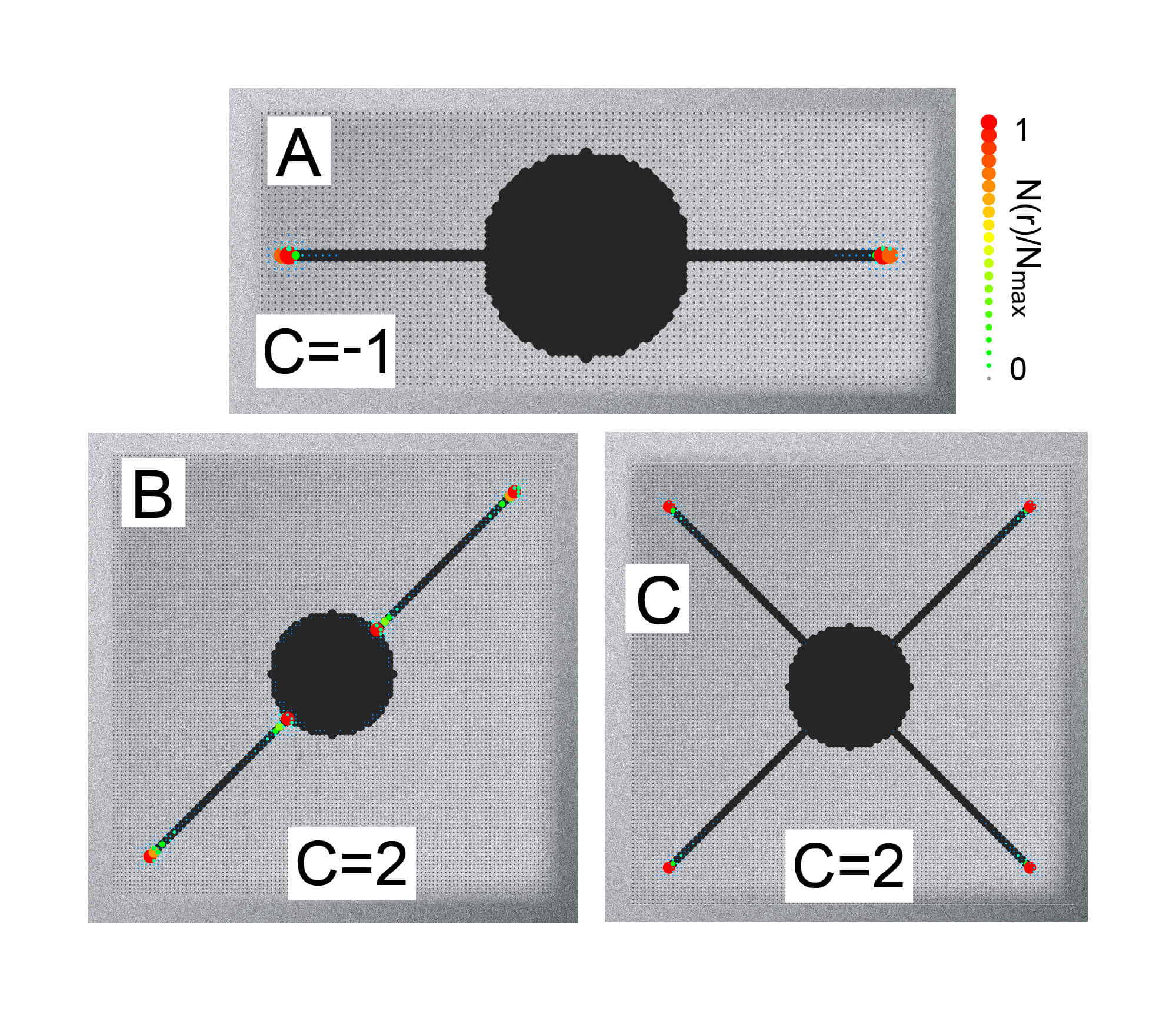}
\caption{\textbf{Real Space counting of the Chern number} {\bf (A)} Shiba island (with $C=-1$ and same parameters as in Fig.~\ref{fig:fig1}) and 2 attached Shiba chains.  Shiba island (with $C=2$ and parameters  $(\mu,\Delta,\alpha,J)=(-0.5t, 0.7t, 0.45t, 2t)$ and {\bf (B)} 2 and {\bf (C)} 4 attached Shiba chains.}
\label{fig:fig2}
\end{figure}
%%%%%%%%%%%%%%%%%%%%%%%%%%%%%%%%%%%%%%%%%%%%%
To demonstrate that such effects exist in experimentally relevant MSH structures, we consider a model that was recently employed to describe the emergence of topological superconductivity in Fe islands deposited on a Re(0001)-O(2x1) surface \,\cite{palacio-morales18arXiv}. Attaching a chain of magnetic atoms to a Fe island [Fig.~\ref{fig:fig1}F], that has the same form as the one realized experimentally\,\cite{palacio-morales18arXiv}, the LDOS shows a zero energy mode localized along the edge of the island, as well as a Majorana fermion localized at the end of the chain. Finally, we note that when a second chain is attached to the Shiba island of [Fig.~\ref{fig:fig1}C], the dispersive Majorana mode moves from the edge of the Shiba island to the end of the second chain and forms a bound state [see Fig.~\ref{fig:fig2}A]. This result suggests  that the localization of Majorana modes at the end of the chains is insensitive to the particular shape of the chain in its middle.\\

Identifying the topological invariant -- the Chern number -- through measurements has been a long sought goal. Attaching chains via atomic manipulation to magnetic islands provides a new approach to detecting the island's Chern number. To demonstrate this, we consider a topological superconductor with Chern number $C=2$, implying that a Shiba island possesses two degenerate chiral Majorana edge modes. When two chains are attached to such an island, one of the Majorana modes is separated into two zero energy Majorana fermions which are located at each of the ends of the two chains, as shown in Fig.~\ref{fig:fig2}B [we note that for this particular set of parameters, the chains are only in a topological phase when they are oriented along the diagonal, but not when they are oriented along the bond directions, which is opposite to the case considered in Fig.~\ref{fig:fig1}]. The second Majorana mode remains localized along the edge of the Shiba island, with large spectral weight concentrated at those points along the edge where the chains are attached. This result is qualitatively different from the $C=1$ case where in the presence of two chains, no low-energy Majorana mode remains along the edge of the Shiba island [see Fig.~\ref{fig:fig2}A]. Only when four chains are attached to the island [Fig.~\ref{fig:fig2}C] we find that four zero-energy Majorana fermions (arising from the two lowest energy Majorana modes of the island) are located at the end of the chains, with no mode remaining along the edge of the island. These results demonstrate a new real space approach to detecting the Chern number of a two-dimensional topological superconductor through atomic manipulation: if spectral weight for a zero-energy state remains located at the edge of the island when $N-1$ chains are attached, but vanishes for $N$ chains, then the Chern number of the 2D topological superconductors is given by $C=N/2$. We emphasize that the described procedure is not limited to small Chern numbers since chains can be added at arbitrary positions of the edge of the original island.\\

\begin{figure}[t!]
\centering
\includegraphics[width=8.6cm]{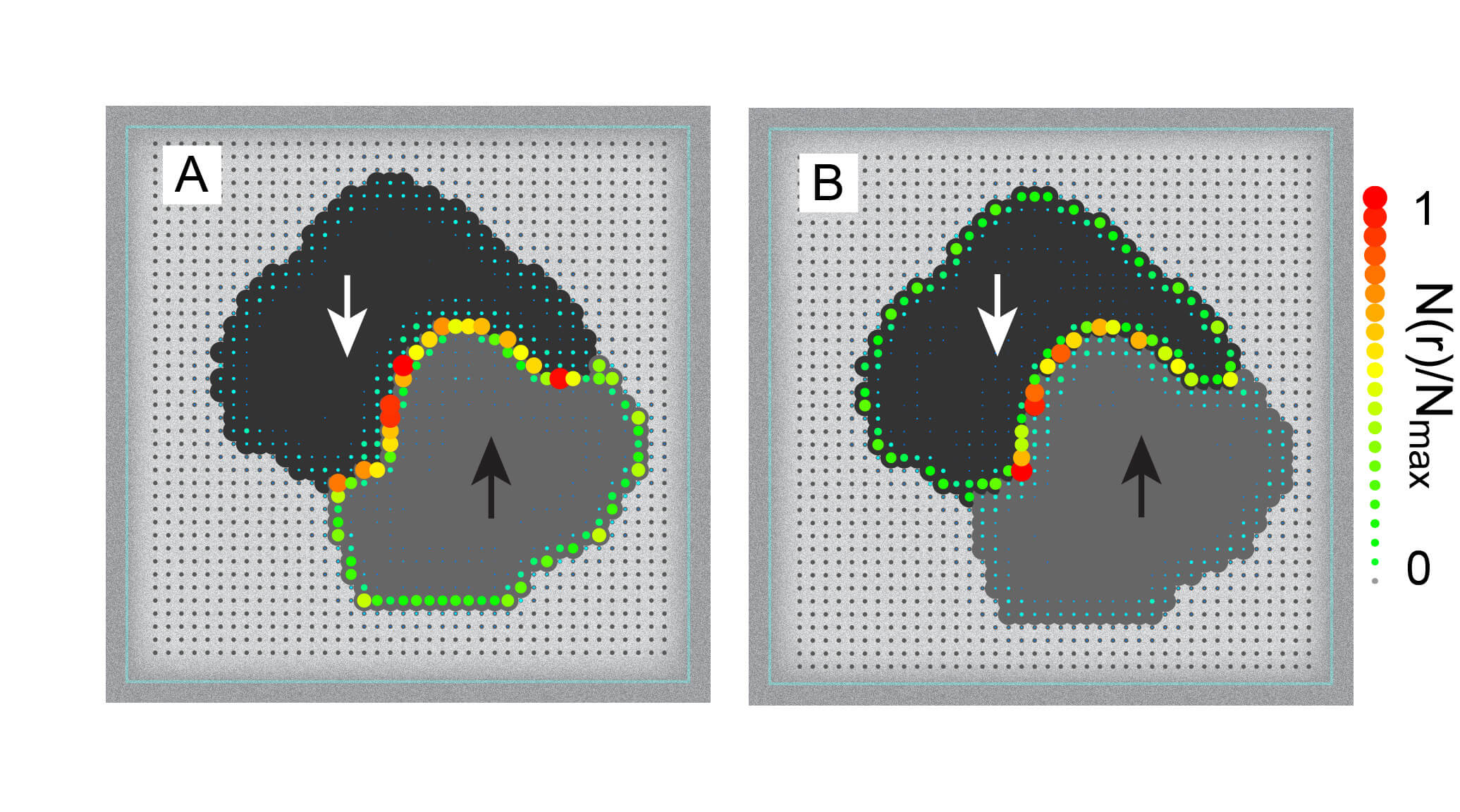}\\[10pt]
\caption{ \textbf{Creation of Majorana modes via magnetic domain walls} MSH structure with 2 magnetic spin-$\uparrow$ (black arrow, gray area) and spin-$\downarrow$ (white arrow, black area) domains and parameters $(\mu,\Delta,\alpha,J)=(-4t, 1.2t, 0.8t, 2t)$. The domains represent topological superconductors with Chern number $C=+ 1$ and $C=- 1$, respectively. {\bf (A)} Spin-$\uparrow$ and {\bf (B)} spin-$\downarrow$ LDOS.
}
\label{fig:fig3}
\end{figure}

The sign of the Chern number characterizing a topological superconductor is determined by the relative alignment of the direction of the magnetic moments, and of the Rashba spin orbit interaction. Thus, two MSH structures with opposite alignment of the magnetic moments possess Chern numbers of opposite sign. This raises the intriguing possibility to create chiral Majorana modes along magnetic domain walls in MSH structures that separate domains of opposite magnetic alignments \cite{Yasuda1311,Wang_PRB024519} and hence with Chern numbers of opposite signs. To demonstrate this, we consider a finite magnetic island consisting of two domains [Fig.~\ref{fig:fig3}] characterized by Chern number $C=\pm 1$. The spin-resolved zero-energy LDOS [Figs.~\ref{fig:fig3}A,B] shows that each of these domains possesses one Majorana mode at the outer boundary, where the Chern number changes from $C=\pm 1$ to $C=0$, and two modes along the (inner) magnetic domain wall, where the Chern number changes from $C=+ 1$ to $C=- 1$. The latter modes are split in real space, and possess a unique spin structure, as revealed in the spin-resolved LDOS [Figs.~\ref{fig:fig3}A,B]. This provides an additional experimental signature of chiral Majorana modes, which could be tested in spin-resolved STM experiments. Moreover, as the ``writing" of magnetic domain walls on the nanoscopic scale has recently been demonstrated using magnetic force microscopy \cite{Yasuda1311}, domain walls provide a unique opportunity to create chiral Majorana modes of arbitrary length, shape and at arbitrary position.\\
%%%%%%%%%%%%%%%%%%%%%%%%%%%%%%%%%%%%%%%%%%%%%
%                  F I G .     3
%%%%%%%%%%%%%%%%%%%%%%%%%%%%%%%%%%%%%%%%%%%%%
%%%%%%%%%%%%%%%%%%%%%%%%%%%%%%%%%%%%%%%%%%%%%

%%%%%%%%%%%%%%%%%%%%%%%%%%%%%%%%%%%%%%%%%%%%%
%                  F I G .     4
%%%%%%%%%%%%%%%%%%%%%%%%%%%%%%%%%%%%%%%%%%%%%
\begin{figure*}[t]
\centering
\includegraphics[width=16.2cm]{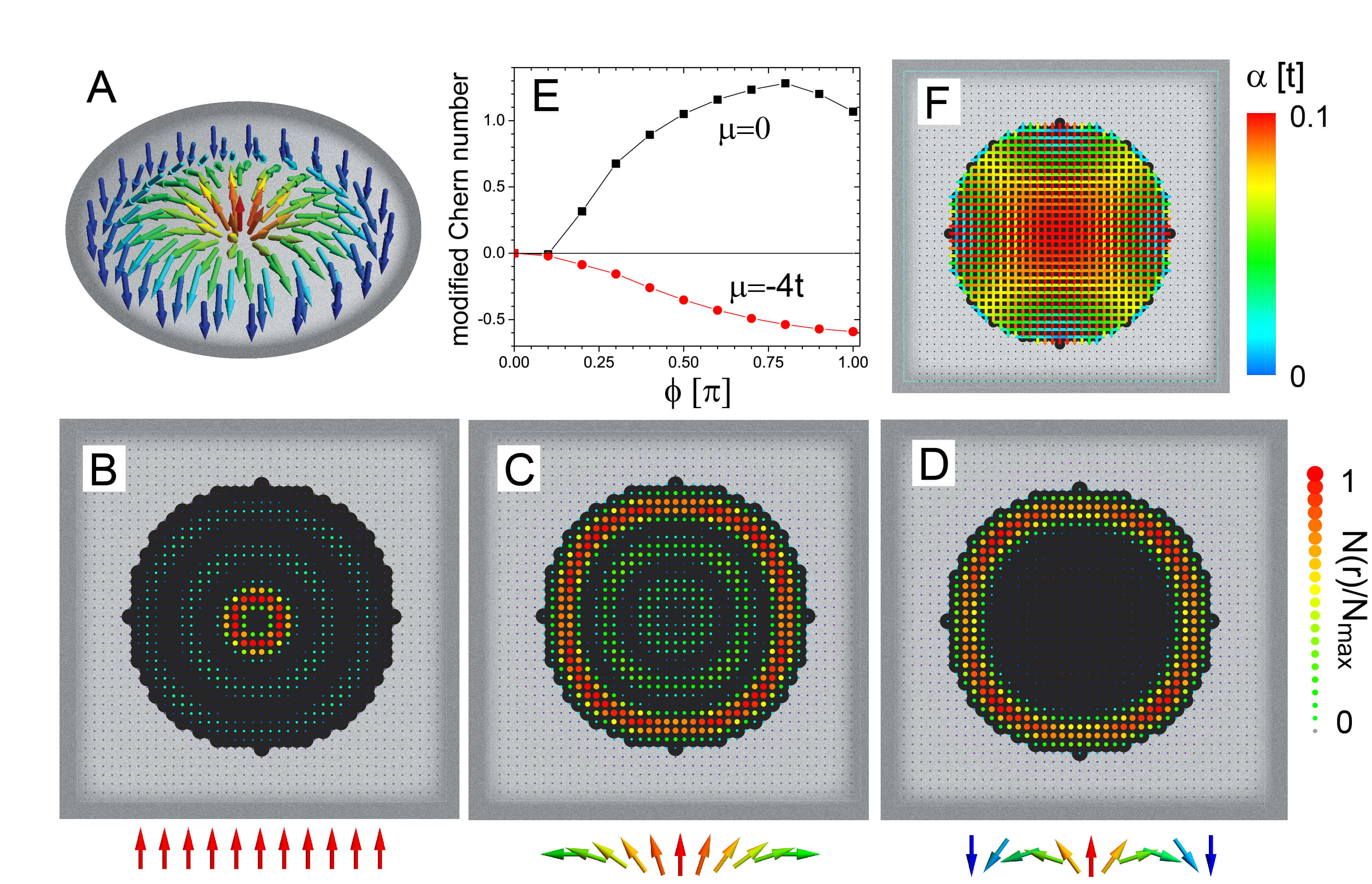}
\caption{\textbf{Creation of Topological Phase through Skyrmions} {\bf (A)} Schematic picture of a skyrmion. Evolution of the zero-energy LDOS with increasing $\Phi$ for a Shiba island of radius $R=15 a_0$ and parameters $(\mu,\Delta,J)=(-4t, t, 1.1t)$ and rotations angles {\bf (B)} $\Phi=0$, {\bf (C)} $\Phi=\pi/2$, and {\bf (D)} $\Phi=\pi$. The lower panels in {\bf (B) - (D)} show schematically the magnetic structure along a cut through the island. {\bf (E)} Modified Chern number (as introduced in Ref.~\cite{RachelMorr17}) as a function of $\Phi$.  {\bf (F)} Magnitude of the Rashba spin-orbit interaction induced by the skyrmion ($\Phi=\pi$).}
\label{fig:fig4}
\end{figure*}
%%%%%%%%%%%%%%%%%%%%%%%%%%%%%%%%%%%%%%%%%%%%%

By imposing a helical magnetic structure onto the Shiba chains in 1D MSH structures, topological superconductivity can be induced even in the absence of Rashba spin-orbit coupling \,\cite{nadj-perge-13prb020407}. The question thus naturally arises to what extent topological superconductivity can arise in 2D Shiba islands by imposing complex magnetic structures. Of particular interest are here magnetic skyrmions \cite{Guang_224505} [Fig.~\ref{fig:fig4}A], as they are not only topological in nature by themselves, but can also be written and deleted using STM techniques \,\cite{romming-13s636,romming-15prl177203}.

To investigate this possibility, we begin by considering a circular Shiba island with a ferromagnetic alignment of spins [Fig.~\ref{fig:fig4}B] which is in a topologically trivial metallic state due to the absence of a Rashba spin-orbit interaction. Such a island possesses a considerable zero-energy LDOS in its center [see Fig.~\ref{fig:fig4}B], with no sign of any edge mode. The spatial structure of the LDOS, however, exhibits a significant evolution when a skyrmion-like magnetic structure is imposed on the Shiba island, as described by the total spin-rotation angle $\Phi$ between the center and the edge of the island [a rotation angle of $\Phi=\pi$ ($\Phi=0$) corresponds  to a magnetic skyrmion (ferromagnet)]. In particular, increasing the radial rotation angle to $\Phi=\pi/2$ [Fig.~\ref{fig:fig4}C] leads to a transfer of spectral weight in the LDOS from the center of the island to the edge, resulting in a well-defined zero-energy edge mode for $\Phi=\pi$ [Fig.~\ref{fig:fig4}D], corresponding to a magnetic skyrmion. That this edge mode is a direct signature of the induced topological superconductivity is confirmed by a calculation of the modified Chern number, ${\cal C}$, of the Shiba island [Fig.~\ref{fig:fig4}E] as a function of $\Phi$ for different chemical potentials. ${\cal C}$ begins to deviate from zero for already rather small values of $\Phi$, implying that the system can be tuned through a topological phase transition by increasing the rotation angle. For a skyrmion magnetic structure ($\Phi=\pi$), the non-zero value of ${\cal C}$ implies that the island is topological in nature, and that the edge mode [see Fig.~\ref{fig:fig4}D] represents a chiral Majorana mode. Note that due to the finite size of the island, as well as the non-collinear magnetic order, ${\cal C}$ is not quantized any longer. The underlying reason for the emergence of topological superconductivity with increasing rotation angle is that such a magnetic structure induces an effective Rashba spin-orbit interaction, as shown in Fig.~\ref{fig:fig4}F for a skyrmion ($\Phi=\pi$) \cite{Mascot_2018}. These results exemplify the creation of a {\it topological Majorana switch}: by writing or deleting skyrmions using STM techniques, it is possible to turn off and on topological superconductivity in a Shiba island.

Our results have demonstrated that the combination of scanning tunneling microscopy and atomic manipulation provides a powerful tool for the atomic scale engineering of Majorana modes in MSH structures. Not only does it provide the ability to tune the dimensionality of Majorana modes, but it also enables the creation of {\it topological switches} by writing and deleting complex magnetic structures, such as skyrmions, in MSH systems. In addition, it establishes a new approach to identifying the topological invariant of the system, the Chern number, in real space. As the experimental techniques necessary for the above-proposed engineering and manipulation of Majorana modes are currently available, they might well hold the key for the realization of the first generation of Majorana-based quantum devices.

%%%%%%%%%%%%%%%%%%%%%%%%%%%%%%%%%%%%%%%%%%%%%
%
%                                                  M E T H O D S
%
%%%%%%%%%%%%%%%%%%%%%%%%%%%%%%%%%%%%%%%%%%%%%
\section{Methods}
We compute the spin-resolved local density of states, $N_\sigma({\bf r})$, from the local, retarded Green's function, $G^r({\bf r,r},\sigma,\omega) $ via
\begin{align}
N_\sigma({\bf r}) = -\frac{1}{\pi} {\rm Im} [ G^r({\bf r,r},\sigma,\omega) ]
\end{align}
$G^r({\bf r,r},\sigma,\omega)$ is obtained from a real-space formalism as described in Ref.\cite{RachelMorr17}. The modified Chern numbers are calculated using a real-space approach as introduced recently \,\cite{RachelMorr17,prodan11jpa113001,prodan17}. The modified Chern number is obtained from the actual Chern number by scaling the latter with the area covered by the Shiba island.

%%%%%%%%%%%%%%%%%%%%%%%%%%%%%%%%%%%%%%%%%%%%%%%%%%%%

\section{Acknowledgements}
The authors would like to thank H.\ Kim, A. Kubetzka, T. Posske, K. von Bergmann, M.\ Vojta and R.\ Wiesendanger for stimulating discussions.
This work was supported by the U. S. Department of Energy, Office of Science, Basic Energy Sciences, under Award No. DE-FG02-05ER46225 (EM, SC, and DKM) and through an ARC Future Fellowship (FT180100211) (SR).

%\bibliography{references_V1}

\end{document}